\documentclass[english,aps,pra,twocolumn,showpacs]{revtex4}
\begin{document}

\title{A Composite  Fermion Approach to the Ultracold Dilute Fermi Gas}
\author{M. A. Cazalilla}
\affiliation{Centro de F\'{\i}sica de Materiales CSIC-UPV/EHU. Paseo Manuel de Lardizabal,
5. E-20018 San Sebasti\'an, Spain,}
\affiliation{Donostia International Physics Center (DIPC), Paseo Manuel de Lardizabal, 4,
E-20018, San Sebasti\'an, Spain,}
\affiliation{Institute for Solid State Physics, University of Tokyo, Kashiwa 277-8581, Japan}
\begin{abstract}
It is argued that the recently observed Fermi liquids in strongly interacting
ultracold Fermi gases are adiabatically connected to  a projected Fermi gas. 
This conclusion is reached by constructing a set of Jastrow wavefunctions, following Tan's observations 
on the structure of the physical Hilbert space [Annals of Physics {\bf 323},  2952  (2008)]. 
The Jastrow projection merely implements  the Bethe-Peierls condition 
on the BCS  and Fermi gas wavefunctions.   This procedure provides a simple  picture of the emergence of 
Fermi polarons as composite fermions in the normal state of the  highly polarized gas. It is  also 
shown that the projected BCS  wavefunction can be written as a condensate of pairs of 
composite fermions (or Fermi polarons).  A Hamiltonian for the composite fermions
is  derived.  Within a mean-field theory, it is shown that the ground state and  excitations of this
Hamiltonian are those of a  non-interacting Fermi gas although  they are described by  Jastrow-Slater wavefunctions. 
 \end{abstract}
\date{\today}
\maketitle

\section{Introduction}\label{sec:intro}

 Recently, several groups~\cite{Duke,ENS,Tokyo} have performed accurate measurements of the
thermodynamics in the unitary regime of  an ultracold dilute Fermi gas. 
The measurements have been found  consistent with 
Ho's~\cite{Ho_universal} universal thermodynamics hypothesis. However, 
the group at ENS~\cite{ENS} also found
 that the equation of state of the \emph{unpolarized}  unitary 
 gas can be  fitted using Fermi liquid theory (FLT) 
in a temperature range from $T_c/\mu \simeq 0.32$  to $T/\mu \simeq  0.8$
($T_c$ being the transition temperature to the superfluid state
and $\mu$ the chemical potential). Based on this analysis,  the ENS group has claimed
that the normal state of the unitary Fermi gas is a  \emph{weakly} correlated Fermi liquid~\cite{ENS}. These results
appear to be in contradiction with the recent spectroscopic observations of the JILA group~\cite{JILA}.
The latter  support the existence of a pseudogap regime~\cite{pseudogapth}, in agreement
with a number of theoretical claims~\cite{pseudogapth,Tsuchiya} (but not others~\cite{Punk}).
Besides the unpolarized gas, Fermi liquid behavior has also been predicted~\cite{Lobo}, 
and observed in a much less controversial  way~\cite{spinimbalance,ENS,ENS2},  
in the normal phase of the  highly polarized  
Fermi gas. In a large region
of the phase diagram, this polaron liquid  also seems to behave as a weakly correlated Fermi liquid. 

 Theoretically, it is not  clear at all how a weakly correlated Fermi liquid can emerge in the unitary limit above
 the critical temperature.   At least na\"ively,  this Fermi liquid cannot be adiabatically connected to the
Hartree-Fock solution  at weak coupling. The reason is that the ground state and excitations of the latter are
 Slater determinants. Such wavefunctions do not obey the Bethe-Peierls condition, which,
as emphasized recently by Tan~\cite{Tan} (see also~\cite{Castin}), 
defines the \emph{physical} Hilbert space of the system. Whether this condition can be fully
implemented  within diagrammatic perturbation theory is also unclear, and may be one of the reasons
why some theories fail to reproduce the experiments~\cite{ENS}\footnote{The Bethe-Peierls condition
can be  implemented in few-body correlation functions (such as the pair correlation function)~\cite{Trento}. 
Such a requirement  is  much weaker than directly imposing the condition on the wavefunction. See Sect.~\ref{sec:projBCS}
for an example dealing with the BCS wavefunction.}. 
So far,  several approaches  have attempted to provide a quantitative description of the normal phase
through sophisticated diagrammatic calculations (see \emph{e.g.} Refs.~\cite{pseudogapth,Punk,Tsuchiya,Sheehy,Sheehy2,Nikolic, Son,Svistunov}
and references therein).  Although these analyses yield a wealth of quantitative predictions, in the  present 
author's opinion, the  picture provided by them is usually not  very  transparent.  On the other hand, the 
picture in terms of wavefunctions  provided by variational methods~\cite{Chevy,Combescot2,Punk2} 
as well as fixed-node diffusion Monte Carlo~\cite{VMC,Giorgini}
is physically more transparent. However,  these methods mostly deal with the ground 
state~\footnote{An important exception is the calculation of the dispersion of
on Fermi polaron  in the highly polarized gas at unitarity~\cite{Lobo}}, and therefore
\emph{a priori} they cannot be applied to understand the properties of the normal state 
at finite temperature. 
 
 Thus, it would desirable to have clearer theoretical picture of the emergence of  Fermi liquid behavior,
especially in the crossover region. Here it is argued  that a more  transparent (although at this point less quantitative) 
picture can be obtained  by incorporating from the scratch Tan's  observation~\cite{Tan} about  the  \emph{physical} Hilbert space
of the system.  The latter is defined as the set of states obeying the Bethe-Peierls condition (see Eq.~(\ref{eq:BPC}) below). 
By projecting the BCS wavefunction onto this space by means of a Jastrow factor, it is shown below
that it can be written as a condensate of pairs of composite fermions. These composite fermions 
are related to the Fermi polarons introduced by Lobo et al.~\cite{Lobo} to describe the normal
state of the highly polarized gas. We show that, in general, the composite fermions (or polarons) 
are the elementary excitations of a projected Fermi gas. They correspond to the original fermions dressed by a density cloud of opposite-spin fermions. 
By deriving a Hamiltonian that acts on unprojected states, 
 it is shown that the Jastrow-Slater wavefunctions are  the eigenstates of a mean-field approximation to such
Hamiltonian. Furthermore, it is found that, at the mean-field level, the   polarons 
behave as non-interacting Fermi gas for all values of the scattering length and spin polarization. The
interactions between them are thus described by corrections to the mean-field theory.  
Nevertheless, although it may be coincidental, we argue that this mean-field theory 
may provide a qualitative explanation for the emergence of 
a weakly correlated Fermi liquid in the crossover regime.
Furthermore,  these results also reveal interesting connections of this problem with the theory of the  fractional 
quantum Hall effect and the physics of the Gutzwiller projection of Hubbard models~\cite{Jain,Read,AndersonJain}. 
 
 The rest of the article is organized as follows: In the next section, we briefly review the mounting experimental and theoretical
 evidence  for Fermi liquid behavior in strongly interacting dilute Fermi gases.
  The properties of the  superfluid grond state are discussed in Sect~\ref{sec:projBCS}  in terms of a projected
 BCS wavefunction.  We use the structure of this wavefunction to motivate the introduction of a set of composite Fermi fields, which
 as shown in Sect.~\ref{sec:projN} are related to the Fermi polarons. In the same section, 
the normal state is considered and we argue that it has the form of a Jastrow-Slater determinant. 
In Sect.~\ref{sec:CFHam}, this state is derived from a  mean-field theory. 
Finally, in section~\ref{sec:concl} we discuss some possible 
extensions to this work and provide our conclusions.

\section{Brief Review of the Relevant Experiments and Theory}

 The findings of the ENS group described in the Introduction  appear to be in  contradiction with  
 the recent angle-resolved photoemission spectroscopy measurements (ARPES)  of JILA group~\cite{JILA}.
Performing ARPES~\cite{ARPES}  above $T_c$ and near unitarity, this group 
observed a prominent BCS-like feature in the spectral function, which 
shows a downturn consistent with a large gap in the single-particle excitation spectrum, 
for $k  \simeq k_F$ ($k_F$  is the Fermi wavenumber). This was observed
at  temperatures where essentially no condensed pairs pairs were found 
using a different technique.   These experimental  results, together with some numerical evidence~\cite{Drummond}, 
come in support of the theories~\cite{pseudogapth,Tsuchiya} claiming that the system 
enters a pseudogap regime above $T_c$.
According to these theories,  such a regime is characterized by the existence of  
preformed Cooper pairs, which have not yet undergone Bose-Einstein 
condensation. Because the electrons remain paired even above $T_c$, 
the density of states at the Fermi surface should be strongly suppressed. 
Therefore, it is  at least na\"ively expected that  the existence of the preformed 
pairs should be also reflected in the thermodynamics as deviations from 
FLT~\cite{Drut}, as observed in the hole-doped cuprate materials~\cite{Loram},
which also exhibit a mysterious pseudogap phase~\cite{Randeriaps}.
 
 On the other hand, if the normal phase of the unpolarized 
gas is a \emph{standard} Fermi liquid,  the spectroscopic properties near the Fermi surface 
must be accounted for  by the Landau quasi-particle (LQP) picture.
In the framework of FLT, the existence of LQPs  is  related to the existence of 
a set of (approximate) eigenstates, $|\mathrm{qp} (N+1,\mathbf{k}, \sigma) \rangle$, with
 total particle number $N+1$ and carrying momentum $\hbar \mathbf{k}$ and spin projection $\sigma$, 
such that:
 \begin{equation}
 \lim_{|\mathbf{k}| \to k_F} \left| \langle \Psi_0(N)  | \psi_{\sigma}(\mathbf{k}) |  
 \mathrm{qp} (N+1,\mathbf{k},\sigma) \rangle 
 \right|^2 =  Z 
 \end{equation}
 where $|\Psi_0(N)\rangle$ is the ground state of the system containing $N$ particles,
 $\psi^{\dag}_{\sigma}(\mathbf{k})$ is
 the Fermi operator that creates an atom carrying momentum $\hbar \mathbf{k}$ and
spin $\sigma$, and  the real number
 $Z$ ($0 < Z \leq 1$) is the quasi-particle weight.  Slightly away from  the Fermi surface, 
 the LQPs acquire a finite lifetime, which grows with the minimun of $T^{-2}$ or $(\varepsilon-\mu)^{-2}$, where $\varepsilon$ is
 the quasi-particle excitation energy.  
 
 The ENS group found the effective mass of the quasi-particles ($m^*$) in the upolarized gas to be
 close to the \emph{bare} atomic mass ($m$), $m^* \simeq 1.13\, m$. Assuming that we are dealing with LQPs and 
 taking into account that the quasi-particle weight $Z \sim m/m^*$, it follows that $Z\sim 1$.
Thus, the density of states at the Fermi energy, which is proportional to $Z$, 
should be rather close to that of the non-interaccting gas in spite of the strong interactions
characteristic of the unitary regime. In the case of $^{3}$He, which is not a dilute
system but where the atoms also interact via a short range potential,  the LQP effective mass  is such that
$2.9 \leq m^{*}/m < 5.7$, depending on the pressure.
Therefore, when compared to $^{3}$He, the unpolarized gas at unitarity appears 
to be a weakly correlated Fermi liquid.  On the other hand,  the downturn observed 
by  the JILA group  would be consistent with a strong reduction of the density of states near the Fermi surface or pseudogap, 
which seems hard to fit into the picture put forward by the ENS group.

 Some clues to better understand the puzzling experimental situation described above  
may come from recent  lattice Monte Carlo calculations~\cite{Drut}. This method (unlike diffusion Monte Carlo)
can access finite temperatures, although calculations become more cumbersome as $T = 0$ is approached. 
Overcoming these difficulties,  Bulgac and coworkers~\cite{Drut} computed
the thermodynamic properties  of the gas as a function of the temperature  
in the crossover regime (for $-2 < k_F a_s < 0.5$, $a_s$ being
the s-wave scattering length)
 down to $T \simeq 0.1 \, T_F$ (where $T_F$ is the Fermi temperature 
of a non-interacting  gas of the same density). 
At unitarity, it was found   that  the unpolarized unitary gas  exhibits free Fermi gas-like behavior
 for $T \gtrsim 0.23 T_F$. Moreover, in a narrow range, between $T_c \lesssim T \lesssim 0.23 T_F$ ($T_c \simeq 0.15 T_F$),
 the system behaves neither as a Fermi liquid nor as a superfluid. Below $T_c$, the energy 
behaves as that of a superfluid, which is accounted for by the joint contributions 
of the Anderson-Bogoliubov mode and the Bogoliubov quasi-particles. 
These numerical results  are in  good agreement with the data obtained by the ENS group
for $T/\mu \gtrsim 0.4$~\cite{ENS},  which are also the most accurately fitted by FLT. Therefore, the results are also consistent with
the scenario (in qualitative agreement with some diagrammatic approaches~\cite{Punk,Tsuchiya})  where the pseudogap regime  exists 
(if at all) only in a temperature window that is rather narrow even at unitarity (but may become broader
on the BEC side of the crossover).  Above the upper limit of this window, 
the system would exhibit FLT-like thermodynamics.

In the case of a strongly polarized Fermi gas at unitarity, the  Fermi liquid state
can be understood as a liquid of Fermi `polarons'. The latter are the result of dressing  the fermions of the minority (say, $\uparrow$) 
spin component   with a cloud of fermions in the majority spin ($\downarrow$) component. At unitarity, it is found~\cite{Lobo,Chevy,Combescot,Punk2} 
that the Fermi polaron has the following dispersion $\epsilon^{p}_{\uparrow}(\mathbf{k}) = E_{b} + \frac{\hbar^2 \mathbf{k}^2}{2m^*}$, 
where $m^*\simeq 1.17 \, m$ and $E_{b}/E_{F \downarrow} \simeq -0.64$, where $E_{F\downarrow}$ 
is the Fermi energy of the majority component,  as determined experimentally in Ref.~\cite{spinimbalance}
 ($m^* \simeq 1.20 \, m$ according to~\cite{ENS}).  In this case again, the effective mass of the polaron
seems to be rather close to the free atom mass, in spite of the strong attractive interactions at unitarity. Further calculations 
using a T-matrix approximation also found small deviations  ( $\sim 50\%$ at most) 
from the bare mass for $-1< (k_Fa_s)^{-1} \lesssim 0.5$ ~\cite{Combescot2}.
Moreover, experimentally the ENS group  found that the interactions between the polarons 
in the polarized gas appear to be negligible at unitarity~\cite{ENS}.
 
 \section{The superfluid state}\label{sec:projBCS}
 
 In the theory of the BCS-BEC crossover, it is believed that the major features 
of  the ground state are captured by the BCS wavefunction~\cite{LeggettBCSBEC,Trento,Castin}:
\begin{eqnarray}
|\Phi_\mathrm{BCS} \rangle &=& \left[ \sum_{\mathbf{k}} \varphi(\mathbf{k})
 \psi^{\dag}_{\uparrow}(\mathbf{k}) \psi^{\dag}_{\downarrow}(-\mathbf{k})\right]^{N/2}  \label{eq:unprojBCS1}\\ 
&=& \left[ \int d\mathbf{x}  d\mathbf{y} \, \varphi(\mathbf{x}-\mathbf{y}) 
  \psi^{\dag}_{\uparrow}(\mathbf{x}) \psi^{\dag}_{\downarrow}(\mathbf{y}) \right]^{N/2} | 0 \rangle. \label{eq:unprojBCS}
\end{eqnarray}
According to this theory, on the BCS side of the crossover (where the scattering length
$a_s < 0$), atoms form Cooper pairs described by the pair
 wavefunction $\varphi(\mathbf{r}) = \frac{1}{\Omega}\sum_{\mathbf{k}} \varphi(\mathbf{k})\:
e^{i \mathbf{k}\cdot \mathbf{r}}$ (here $\Omega$ is the volume of the system).
The pairs form a Bose-Einstein condensate in which the center of mass state at
$\mathbf{k} = \mathbf{0}$ is macroscopically ocupied~\footnote{Indeed, this feature, namely 
the lack of finite momentum pairs
in  $|\Phi_\mathrm{BCS}\rangle$, makes it unsuitable to describe collective excitations such as the Anderson-Bogoliubov
mode, which is the Goldstone mode appropriate to a neutral superfluid.  However, we shall be concerned here 
with the fate of individual  fermionic excitations, and therefore the BCS wavefunction will be sufficient to the purpose of our 
discussion}. 
As the scattering length $a_s$ is tuned by means of a Feshbach resonance towards unitarity 
(where $k_F a_s \to - \infty$),
the size of the pairs shrinks  until a two-body
bound state (a molecular dimer) forms. Further on,  as $a_s$ continues to grow positive, the dimers
become more tightly bound on  the BEC side of the crossover.
 
  In an important development in the theory of the BCS-BEC crossover, 
Tan~\cite{Tan} has  recently 
emphasized (see also~\cite{Castin}) the importance 
of the Bethe-Peierls condition (BPC) in defining the \emph{physical} Hilbert space of a \emph{dilute} 
ultracold Fermi gas. This condition states that the  
$s$-wave interactions, which are  also described in terms of the Lee-Huang-Yang pseudo-potential~\cite{LeeYangHuang} $V_\mathrm{int}(\mathbf{r}) = 
\frac{4\pi \hbar^2 a_s}{m}  \delta (\mathbf{r}) \partial_{r}\left(r \cdot \right)$, can be
replaced by a boundary condition in the limit where the range of the potential goes to zero (\emph{i.e.} $r_0 \to 0$).
For the many-particle states, this means that, provided 
three and higher-body interactions can be neglected (that is, the gas is dilute and Efimov bound states
are not present),  any  physical state  of the system containing $N_{\uparrow}$ spin-up fermions
and $N_{\downarrow}$ spin-down fermions exhibits the following behavior:
\begin{eqnarray}
\lim_{|\mathbf{x}_i -\mathbf{y}_j|\to 0}  \Psi_\mathrm{phys}(\left\{\mathbf{x}_k\right\}_{k=1}^{N_{\uparrow}};
\left\{\mathbf{y}_l\right\}_{l=1}^{N_{\downarrow}}) = \left( \frac{1}{|\mathbf{x}_i -\mathbf{y}_j|} - \frac{1}{a}\right) \nonumber \\
\times \tilde{\Psi}(\frac{\mathbf{x}_i + \mathbf{y}_j}{2},\left\{\mathbf{x}_k\right\}_{k\neq i},  \left\{\mathbf{y}_l\right\}_{l\neq j}), 
\quad \quad
\label{eq:BPC}
\end{eqnarray}
for any pair of opposite spin particles ($i=1,\ldots, N_{\uparrow}$ and $j=1,\ldots, N_{\downarrow}$).   
As Tan discovered, the BPC has important consequences for the short-distance correlation functions and
the total energy of the states. In particular, he predicted~\cite{Tan} that the momentum distribution of any physical
state $\Psi$ behaves as $C_{\Psi} \: k^{-4}$ at large wavenumber, $k$, where  Tan's contact $C_{\Psi} = \langle \Psi | \rho_{\uparrow}(\mathbf{r}) \rho_{\downarrow}(\mathbf{r}) | \Psi \rangle$. This prediction has been recently verified by the JILA group~\cite{JILA_Tan}.   

 Tan also pointed out~\cite{Tan} that the BCS wavefunction, Eqs.~(\ref{eq:unprojBCS1},\ref{eq:unprojBCS}), 
does not satisfy the BPC and therefore  it is 
not in the physical space. This is so  even if  the behavior of the pair wavefunction, $\varphi(\mathbf{r})$, 
matches the solution of the two-body problem as $|\mathbf{r}| \to 0$. 
Indeed, by using the determinantal representation of the BCS wavefunction (cf. Eq.~\ref{eq:bcsdet}),
it can be shown that  the many-body wavefunction constructed from such a $\varphi(\mathbf{r})$ 
cannot be written as in Eq.~(\ref{eq:BPC}), where $\tilde{\Psi}$ depends only 
on $(\mathbf{x}_i+\mathbf{y}_j)/2$  and the coordinates of the other particles but not 
on $\mathbf{x}_i -\mathbf{y}_j$. The BPC is also often imposed on the anomalous
correlator~\cite{Trento},   by requiring that
\begin{equation}
\langle \psi_{\uparrow}(\mathbf{k})\psi_{\downarrow}(-\mathbf{k})\rangle 
\sim \frac{4\pi}{k^2}, 
\end{equation}
for $k \gg k_F$ ($ \frac{4\pi}{k^2}$ is the Fourier transform of the term $\frac{1}{r}$ in the BPC). This is akin to imposing the
condition on an expectation value, which is a necessary but not sufficient requirement for 
Eq.~(\ref{eq:BPC}) to hold. In fact, when the expectation
value is taken over the BCS state, this is equivalent to imposing the condition on the pair wavefunction $\varphi(\mathbf{k})$
because $\langle \psi_{\uparrow}(\mathbf{k})\psi_{\downarrow}(-\mathbf{k})\rangle = \varphi(\mathbf{k})/(1+ |\varphi(\mathbf{k})|^2)$. 

  In spite of the shortcoming pointed by Tan,  
the BCS wavefunction  still captures the long-distance correlations of particles in the ground state, as discussed above. 
In a sense,  it is the simplest ansatz that describes a condensate of fermion pairs 
that turn into molecular dimers as the interaction is tuned across the
Feshbach resonance. Therefore, we shall adopt the simple-minded 
approach that the shortcoming of BCS wavefunction  
can be fixed by projecting it onto the physical  Hilbert space. Indeed, it has been recently shown~\cite{MoraWaintal}   that, 
in the quantum Monte Carlo method, a good measure of the quality of a trial wavefunction is its overlap with the ground state. Since the latter
is certainly in the physical Hilbert space, projection simply gets rid of the undesired components of the trial wavefunction that 
decrease the overlap and increase the variational energy.  Thus, on a formal level, 
we can define a projector, $\mathcal{J}_{\mathrm{BPC}}(a_s)$, onto the space of 
states obeying Eq.~(\ref{eq:BPC}) for a given value of the scattering length, 
$a_s$, so that, a variational sense, we choose to work with:
\begin{equation}
|\Psi_\mathrm{BCS}\rangle = \mathcal{J}_{\mathrm{BPC}}(a_s) |\Phi_\mathrm{BCS}\rangle. \label{eq:projectBCS}
\end{equation}
Practically, the projection can be implemented in various ways, and the result should be
independent of the details  provided it is optimal in 
a variational sense. One method that is particularly popular in 
fixed-node Monte Carlo calculations is a Jastrow factor~\cite{VMC,Giorgini}. Following those
approaches, we write the wavefunction as
(in what follows we collectively denote by
$\mathbf{X} = \left\{\mathbf{x}_k\right\}$  the coordinates of the spin-up fermions 
and by  $\mathbf{Y} = \left\{\mathbf{y}_l\right\}$ those of the spin-down fermions):
\begin{eqnarray}
\Psi_\mathrm{BCS}(\mathbf{X}, \mathbf{Y}) &=& \prod_{i, j=1}^{N/2} \chi_J(\mathbf{x}_i - \mathbf{y}_j) 
\: \Phi_\mathrm{BCS}(\mathbf{X}, \mathbf{Y})  \label{eq:projBCS}, \\
\Phi_\mathrm{BCS}(\mathbf{X}, \mathbf{Y}) &=& \mathrm{det}\left[ \varphi(\mathbf{x}_i - \mathbf{y}_j)\right],\label{eq:bcsdet} 
\end{eqnarray}
where we have written the unprojected BCS function, $\Phi_\mathrm{BCS}$ 
as a determinant~\cite{Bouchaudetal} (for $N_{\uparrow} = N_{\downarrow} = N/2$). The above wavefunction should not be 
regarded as optimal as far as the calculation of  the ground
state energy is concerned, in so
much as $\Phi_\mathrm{BCS}$ is also not the best  (unprojected) variational state and many 
improvements are possible, such additional Jastrow factors, Feynman backflow, etc~\cite{VMC}. 
However, this wavefunction is sufficiently simple for the purpose of discussing the 
change of picture brought about by the projection. In order to fullfil (\ref{eq:BPC}) we demand that
$\chi_J(\mathbf{r}) \sim \left(\frac{1}{r} - \frac{1}{a}  + O(r)\right)$ and $\varphi(\mathbf{r}) = \mathrm{const}.
+ O(r^2)$ for $r_0 \ll r \ll \mathrm{min}\left\{ a_s, k^{-1}_F  \right\}$.  
The Jastrow is only required to correct the short-distance behavior of the wavefunction, and therefore
 $\chi_J(\mathbf{r}) \to 1$ for $r \gg k^{-1}_F$. The detailed behavior 
 of  $\chi_J(\mathbf{r})$  between these two asymptotic limits must be determined 
variationally and it depends on the gas parameter $k_F a_s$  and the spin polarization. 
Moreover, since $\chi_J(\mathbf{r})$ is a  zero-energy solution of the two-body problem
for $r_0 \to 0$ (see next section),  we expect that, deep into the BCS and BEC regime where $|k_F a_s| \lesssim 1$, 
it strongly deviates from unity only in rather small region, $r \lesssim |a_s|$.  

Next we note that, by introducing $J_i(\mathbf{x}_i;\mathbf{X}) = \prod_{j=1}^{N/2} \chi_{J}(\mathbf{x}_i - \mathbf{y}_j)$
 and $J_j(\mathbf{Y};\mathbf{y}_j) =  \prod_{i=1}^{N/2} \chi_{J}(\mathbf{x}_i - \mathbf{y}_j)$, 
 it is possible to write the projected BCS state (\ref{eq:projBCS}) as a determinant:
\begin{eqnarray}
\Psi_\mathrm{BCS}(\mathbf{X},\mathbf{Y}) = \det\left[ \varphi(\mathbf{x}_i -\mathbf{y}_j) J_i(\mathbf{x}_i;\mathbf{Y}) \right] \\
= \det\left[ \varphi(\mathbf{x}_i -\mathbf{y}_j) J_j(\mathbf{X};\mathbf{y}_j) \right]\\
=   \det\left[ \varphi(\mathbf{x}_i -\mathbf{y}_j) J^{1/2}_i(\mathbf{x}_i;\mathbf{Y}) J^{1/2}_j(\mathbf{X};\mathbf{y}_j) \right]
\end{eqnarray}
Therefore, we can regard the above wavefunction again as a BCS state, not of the original 
fermions, but  of a system of `composite fermions' instead, which are defined by attaching to each
fermion a Jastrow factor.  This can be seen more clearly by working in second quantization. By analogy 
to  the theory of composite particles in the fractional quantum Hall effect~\cite{Read,Sondhi}, 
we introduce two (hermitian) operators $U_{\sigma}(\mathbf{r})$ ($\sigma=\uparrow,
\downarrow$)  having the  following property $U_{\sigma}(\mathbf{r}) \psi^{\dag}_{-\sigma}(\mathbf{s}) = 
\chi_J(\mathbf{r}-\mathbf{s}) \psi^{\dag}_{-\sigma}(\mathbf{s})U_{\sigma}(\mathbf{r})$, but otherwise commuting.  
Indeed,  provided that $\chi_J(\mathbf{r})$ is a positive function, an explicit construction of  these operators is
$U_{\sigma}(\mathbf{r}) = \exp\left[\int d\mathbf{r}' \, \ln \chi_J\left(\mathbf{r}-\mathbf{r}'\right) \rho_{-\sigma}(\mathbf{r}') \right]$.  Thus, if we define the following quasi-particle 
operator $\pi^{\dag}_{\sigma}(\mathbf{r}) = \psi^{\dag}_{\sigma}(\mathbf{r}) U_{\sigma}(\mathbf{r})$,
the above state can be written as a condensate of pairs of these quasi-particles:
\begin{eqnarray}
|\Psi_\mathrm{BCS} \rangle &=& \left[\int d\mathbf{x} d\mathbf{y} \, 
\varphi(\mathbf{x}-\mathbf{y}) \pi^{\dag}_{\uparrow}(\mathbf{x})
\pi^{\dag}_{\downarrow}(\mathbf{y}) \right]^{N/2} |0\rangle,\quad \\
&=& \left[ \sum_{\mathbf{k}} \varphi(\mathbf{k}) \pi^{\dag}_{\uparrow}(\mathbf{k}) \pi^{\dag}_{\downarrow}(\mathbf{-k})\right]^{N/2} |0\rangle.
\end{eqnarray}
It is also possible  to define a quasi-hole operator $\eta_{\sigma}(\mathbf{r}) = U^{-1}_{\sigma}(\mathbf{r})\psi_{\sigma}(\mathbf{r})$.
It can be shown that $\left\{ \pi^{\dag}_{\sigma}(\mathbf{r}), \pi^{\dag}_{\sigma'}(\mathbf{r}') \right\} = \left\{ \eta_{\sigma}(\mathbf{r}),
\eta_{\sigma'}(\mathbf{r}')\right\} = 0$ and $\left\{ \eta_{\sigma}(\mathbf{r}), 
\pi^{\dag}_{\sigma'}(\mathbf{r}') \right\} = \delta_{\sigma,\sigma'}\delta(\mathbf{r}-\mathbf{r}')$.  However, 
note that  $\pi^{\dag}_{\sigma}(\mathbf{r}) \neq \left[\eta_{\sigma}(\mathbf{r})\right]^{\dag}$. Therefore,
these operators should be handled with care as they result from a non-unitary transformation and the states
created by them may turn out to be non-orthogonal (see Sect.~\ref{sec:projN} for further discussion).

  The wavefunction in~(\ref{eq:projBCS}) can be regarded as a simple model for the ground 
 state of the superfluid state. The Jastrow factor introduces  attractive correlations between fermions in different
pairs  because, at short distances,   $\chi_J(\mathbf{r})$ matches the solution of the two-body problem which describes two-particle attraction 
for $a_s < 0$ and a shallow bound state for $a_s  > 0$.  These correlations can be regarded  as a `pairing frustration' mechanism. 
The mechanism is less effective deep into the BCS and BEC regimes, where $\chi_J(\mathbf{r})$ 
is short-ranged in the sense that it strongly deviates from unity only for $r \lesssim |a_s|$. 
Therefore,  the projected wave function will have a large overlap (at finite $N$) with the unprojected BCS wavefunction. 
Nevertheless, in the deep BEC regime, the short range Jastrow correlations describe the interactions between the molecular dimers. 
On the other hand,
in the crossover regime where $|k_F a_s| \gg 1$, the size of the pairs and the range of the Jastrow
are both comparable to $k^{-1}_F$, the mean inter-particle distance. Therefore, pairing will be strongly frustrated, 
which will lead to a reduction of the superfluid density from the unprojected BCS state.  As to  the fermionic excitations (Bogoliubov
particles), they can be obtained by projecting  the Bogoliubov excitations of the unprojected state $|\Phi_\mathrm{BCS}\rangle$.
However, since we are interested in the normal state, we shall
not pursue this analysis here. 
   
\section{The normal state}\label{sec:projN}
 
  When obtained by variationally minimizing the energy of $|\Psi_{BCS} \rangle = \mathcal{J}_\mathrm{BPC}(a_s) | \Phi_\mathrm{BPC}\rangle$,
the gap  $\Delta$ that parametrizes the unprojected BCS state, $|\Phi_\mathrm{BCS}\rangle$, 
is a function of $a_s$, the mean  density,  and the spin polarization measured
by \emph{e.g.} $P = |N_{\uparrow} - N_{\downarrow}|$.  Nevertheless,  
let us for a while consider $\Delta$ as independent of $a_s$  and imagine that
the gap collapses  (\emph{i.e.} $\Delta \to 0$)  while keeping $a_s$ constant. Physically, the collapse can be
caused either by thermal fluctuations or
by making $P$ sufficiently large~\footnote{If the gap collapses due to thermal fluctuations,
the state of the system is a mixed state. However, we adopt the point of view that such a mixed state can be 
constructed from excitations of the normal state. It is also not implied
the gap will collapse in a continuous transition as the spin polarization grows.}. 
Na\"ively, the state that results from such a limit would be 
$|\Phi_\mathrm{FS}\rangle = \prod_{|\mathbf{k}| < k_{F\uparrow}} 
\psi^{\dag}_{\mathbf{k}\uparrow}    \prod_{|\mathbf{k}| < k_{F\downarrow}}  \psi^{\dag}_{\mathbf{k}\downarrow}|0\rangle$. 
However,  this state is unphysical because it 
does \emph{not} satisfy the BPC,  Eq.~(\ref{eq:BPC}), for the given value of $a_s$.  
The correct state  is $|\Psi_\mathrm{N}\rangle = \mathcal{J}_\mathrm{BPC}(a_s) |\Phi_\mathrm{FS}\rangle$.  In terms of the 
quasi-particle operators, $\pi^{\dag}_{\sigma}(\mathbf{k})$, the projected normal state reads:
\begin{equation}
|\Psi_\mathrm{N} \rangle = \prod_{|\mathbf{k}| < k_{F\uparrow}} \pi^{\dag}_{\uparrow}(\mathbf{k})
\prod_{|\mathbf{k}| < k_{F\downarrow}}   \pi^{\dag}_{\downarrow}(\mathbf{k}) |0\rangle.
\end{equation}
In coordinate representation this state is the following Jastrow-Slater 
wavefunction ($|\mathbf{k}_{\alpha}|<k_{F\uparrow}, |\mathbf{k}_{\beta}|<k_{F\downarrow}$):
\begin{equation}
\Psi_\mathrm{N} = \prod_{i,j} \chi_J(\mathbf{x}_i-\mathbf{y}_j)\:
 \mathrm{det}\left[ \phi_{\mathbf{k}_{\alpha}}(\mathbf{x}_{i}) \right]
\mathrm{det}\left[ \phi_{\mathbf{k}_{\beta}}(\mathbf{y}_{j}) \right]. \label{eq:JS}
\end{equation}
From this form, it can be shown that $\pi^{\dag}_{\sigma}(\mathbf{k}) |\Psi_{\mathrm{N}}\rangle = 0$ if $|\mathbf{k}| < k_{F\sigma}$ and 
$\eta_{\sigma}(\mathbf{k})  |\Psi_{\mathrm{N}}\rangle = 0$ if $|\mathbf{k}| >  k_{F\sigma}$. In words, the state has
a well defined Fermi surface.

  In order to obtain a better insight into the excitations created by 
 the quasi-particle operator $\pi_{\sigma}(\mathbf{k})$, let us  focus on a state describing one spin-up fermion 
 'impurity' in a Fermi sea of $N_{\downarrow}$ fermions \footnote{Note that projection onto the physical 
space is not needed for fully polarized states.}:
\begin{eqnarray}
\langle \mathbf{x}, \mathbf{Y}| \pi^{\dag}_{\uparrow}(\mathbf{x})  \prod_{|\mathbf{k}_{\alpha}| <k_{F\downarrow}} \pi^{\dag}_{\downarrow}(\mathbf{k}_{\alpha}) | 0  \rangle = 
\langle \mathbf{x}, \mathbf{Y}| \psi^{\dag}_{\uparrow}(\mathbf{x})  U_{\uparrow}(\mathbf{x}) \nonumber \\ 
\times \prod_{|\mathbf{k}_{\alpha}| <k_{F\downarrow}}   \psi^{\dag}_{\downarrow}(\mathbf{k}_{\alpha}) | 0  \rangle 
= \prod_{j=1}^{N_{\downarrow}} \chi_J(\mathbf{x}-\mathbf{y}_i) \mathrm{det}\left[ \phi_{\mathbf{k}_{\alpha}}(\mathbf{y}_j) \right], \quad
\end{eqnarray}
where $\phi_{\mathbf{k}}(\mathbf{r}) = \Omega^{-1/2} e^{i \mathbf{k}\cdot \mathbf{r}}$ and $\langle \mathbf{x},\mathbf{Y}| = 
\langle 0| \psi_{\uparrow}(\mathbf{x}) \psi_{\downarrow}(\mathbf{y}_1)\cdots\psi_{\downarrow}(\mathbf{y}_M)$. The above state describes
 a `Fermi polaron' and it was first introduced in the analysis of the normal state of the highly spin-polarized Fermi gas at unitarity~\cite{Lobo}.
Indeed, by expanding the exponent of $U_{\uparrow}(\mathbf{x}) = \exp\left[ \int d\mathbf{z} \ln \chi_J(\mathbf{x}-\mathbf{z}) \rho_{\downarrow}(\mathbf{z})\right]$ to lowest order, the following state  is obtained ($|\Phi^{\downarrow}_{FS} \rangle =   
\prod_{|\mathbf{k}_{\alpha}| <k_{F\downarrow}} \psi^{\dag}_{\downarrow}(\mathbf{k}_{\alpha}) | 0  \rangle$, 
$| \mathbf{k} (\uparrow) \rangle = \psi^{\dag}_{\uparrow}(\mathbf{k})|0\rangle$, $f_{J}(\mathbf{q}) = \int d\mathbf{s} \: \ln 
\chi_J(\mathbf{s}) \: e^{-i \mathbf{q}\cdot \mathbf{r}}$):
\begin{eqnarray}
|\pi(\mathbf{k},\uparrow) \rangle &\simeq& |\mathbf{k} (\uparrow)\rangle | \Phi^{\downarrow}_\mathrm{FS}\rangle + 
\frac{1}{\Omega}\sum_{\mathbf{p q}}  f_J(\mathbf{q}) 
\: |\mathbf{k}-\mathbf{q} (\uparrow)\rangle \nonumber\\
&& \times  \psi^{\dag}_{\downarrow}(\mathbf{p}+ \mathbf{q}) \psi_{\downarrow}(\mathbf{p}) | \Phi^{\downarrow}_\mathrm{FS}\rangle, 
\end{eqnarray}
which has been employed in simple variational approaches to the problem~\cite{Chevy,Combescot,Punk2}. The above equations show that the composite
fermionic excitations described by $\pi_{\sigma}(\mathbf{k})$ correspond to the original fermions dressed by a cloud of opposite spin fermions. 

 In general, for arbitrary numbers $N_{\uparrow}$ and $N_{\downarrow}$, the actual energy of $|\Psi_{N}\rangle$ 
must be obtained by variationally optimizing the Jastrow projector so that 
\begin{eqnarray}
E^N_0 &=&\frac{\langle \Psi_{N} | H | \Psi_{N}\rangle}{\langle \Psi_{N} | \Psi_{N}\rangle} \\
&=& \frac{\langle \Phi_\mathrm{FS}|\mathcal{J}_\mathrm{BPC}(a_s) H \mathcal{J}_\mathrm{BPC}(a_s)|  \Phi_\mathrm{FS}\rangle}
{\langle \Phi_\mathrm{FS}|\mathcal{J}^2_\mathrm{BPC}(a_s)| \Phi_\mathrm{FS}\rangle} \label{eq:en0}
\end{eqnarray}
is minimum. For $N_{\uparrow} = N_{\downarrow}$, using the Jastrow-Slater state  $\Psi_N$  as the trial
wavefunction of  a fixed-node diffusion Monte Carlo calculation, it was  found in Ref.~\cite{Lobo} 
that the energy per particle is  about $30\%$ higher than  the energy of the  superfluid state. This shows  that  the normal
state has a superfluid instability. The latter can be described as condensation of pairs of 
composite fermions,  according to the discussion in Sect.~\ref{sec:projBCS}.

The excitation energy of a quasi-particle state is obtained by evaluating:
\begin{eqnarray}
&&\epsilon^{p}_{\sigma}(\mathbf{k}) + E^N_0 = 
\frac{ \langle \Psi_{N} | \pi_{\sigma}(\mathbf{k}) H  \pi^{\dag}_{\sigma}(\mathbf{k})|\Psi_{N}\rangle  \nonumber}{\langle \Psi_{N} |\pi_{\sigma}(\mathbf{k})
\pi^{\dag}_{\sigma}(\mathbf{k})|\Psi_{N}\rangle}\\
 &=& \frac{\langle \Phi_\mathrm{FS}|\psi_{\sigma}(\mathbf{k}) \mathcal{J}_\mathrm{BPC}(a_s) H  
\mathcal{J}_\mathrm{BPC}(a_s)\psi^{\dag}_{\sigma}(\mathbf{k}) | \Phi_\mathrm{FS}\rangle}{\langle \Phi_\mathrm{FS}|
\psi_{\sigma}(\mathbf{k}) \mathcal{J}^2_\mathrm{BPC}(a_s) \psi^{\dag}_{\sigma}(\mathbf{k}) | \Phi_\mathrm{FS}\rangle}, 
\quad \label{eq:ekn}
\end{eqnarray} 
(for quasi-hole states, $\pi^{\dag}_{\sigma}(\mathbf{k})$ should be replaced by $\eta_{\sigma}(\mathbf{k})$).
Note that the Jastrow factor in this calculation should be built from the same $\chi_{J}(\mathbf{r})$ function as for 
the ground state. The logic behind such an approximation is that the ground state correlations are not 
strongly modified by the creation of a few low-energy excitations.

\section{Mean Field Theory Approach}~\label{sec:CFHam}
 Although it is possible to deal with the composite fermions (CFs)  numerically using the Jastrow-Slater
 wavefunctions introduced above, it can be also useful to derive an effective Hamiltonian that describes them. 
Indeed, we show in the following that the normal state $|\Psi_\mathrm{N}\rangle$ is the ground state of 
 a certain mean-field (MF) Hamiltonian. Let us first derive the Hamiltonian for the unprojected states
 by considering the  many-body Schr\"odinger equation for the physical states:
 \begin{equation}
 i \hbar 
\partial_t \Psi_{\mathrm{phys}}(\mathbf{X},\mathbf{Y},t)  = H \Psi_{\mathrm{phys}}(\mathbf{X},\mathbf{Y},t) 
\label{eq:schro}
 \end{equation}
 where we make no particular assumption about the form of $\Psi_{\mathrm{phys}}(\mathbf{X},\mathbf{Y},t)$.
 The Hamiltonian,
 \begin{eqnarray}
 H &=&  \sum_{i=1}^{N_{\uparrow}} \frac{\mathbf{p}^2_{i\uparrow}}{2m} +   \sum_{j=1}^{N_{\downarrow}} \frac{\mathbf{p}^2_{j\downarrow}}{2m}
 + \sum_{i,j} V(\mathbf{x}_i - \mathbf{y}_i).
   \label{eq:ham}
 \end{eqnarray} 
  In the above expression $\mathbf{p}_{i\uparrow}$ ($\mathbf{p}_{j\downarrow}$)
 is the momentum operator of the $i$-th ($j$-th) spin-up (spin-down) particle, $\mathbf{x}_i$ ($\mathbf{y}_j$) its 
 position operator,  and $V(\mathbf{x}-\mathbf{y})$ is 
a short-range two-body  potential, which in the limit $r_0 \to 0$ can be
 replaced by the BPC, Eq.~(\ref{eq:BPC}). Upon introducing $\Psi_{\mathrm{phys}}(\mathbf{X},\mathbf{Y},t)
 = \mathcal{J}_\mathrm{BPC}(a_s) \Phi_\mathrm{CF}(\mathbf{X},\mathbf{Y},t) = 
 \prod_{i,j}\chi_J(\mathbf{x}-\mathbf{y}_j) \Phi_\mathrm{CF}(\mathbf{X},\mathbf{Y},t)$ into 
 (\ref{eq:schro}), we arrive at:
 \begin{equation}
 i  \hbar  \mathcal{J}_\mathrm{BPC}(a_s)\partial_{t}\Phi_\mathrm{CF} = 
\mathcal{J}_\mathrm{BPC}(a_s) H_\mathrm{CF} \Phi_\mathrm{CF}, 
 \end{equation}
 where
 \begin{eqnarray}
 H_\mathrm{CF} &=&  \sum_{i=1}^{N_{\uparrow}} \frac{\left(\mathbf{p}_{i\uparrow} + \mathbf{v}_{\uparrow}(\mathbf{x}_i) \right)^2}{2m}
 + \sum_{j=1}^{M_{\downarrow}} \frac{\left(\mathbf{p}_{j\downarrow} + \mathbf{v}_{\downarrow}(\mathbf{y}_j) \right)^2}{2m} \nonumber \\
 &&+ \sum_{i,j} V(\mathbf{x}_i-\mathbf{y}_j),  \label{eq:hamcf}
\end{eqnarray}
the vectors $\mathbf{v}_{\sigma}(\mathbf{r}) =  -i s_\sigma \: \hbar \int d\mathbf{s} \: 
 \frac{\nabla_\mathbf{r} \chi_J(\mathbf{r}-\mathbf{s})}{\chi_J(\mathbf{r}-\mathbf{s})} \: \rho_{-\sigma}(\mathbf{s})$,
where $\rho_{\sigma}(\mathbf{r}) = \sum_{k} \delta(\mathbf{r} - \mathbf{r}_k)$ is the density operator of the
CFs carrying spin $\sigma$ and $s_\uparrow = - s_\downarrow= +1$. 
In retrospect, it may seem that we have gained very little by the previous transformation. The CF 
Hamiltonian, Eq.~(\ref{eq:hamcf}), looks even more complicated that the original one, Eq.~(\ref{eq:ham}).
However, it can be shown that the presence of the interaction potential in (\ref{eq:hamcf}) is somewhat redundant: Upon
expanding the kinetic energy term operator, the following term appears:
\begin{eqnarray}
\sum_{i,j} \frac{1}{\chi_J(\mathbf{x}_i -\mathbf{y}_j)} \Bigg[ -\frac{\hbar^2}{2m} \left(\nabla^2_{\mathbf{x}_i}
+ \nabla^2_{\mathbf{y}_j} \right)   \nonumber\\ 
+ V(\mathbf{x}_i - \mathbf{y}_j) \Bigg] \chi_{J}(\mathbf{x}_i - \mathbf{y}_j),   
\end{eqnarray}
which vanishes because $\chi_J(\mathbf{r})$ obeys:
\begin{equation}
\left[ -\frac{\hbar^2}{m} \nabla^2_{\mathbf{r}} + V(\mathbf{r}) \right] \chi_J(\mathbf{r}) = 0.
\end{equation}
Hence,
\begin{eqnarray}
H_{\mathrm{CF}} &=& \sum_{i=1}^{N_{\uparrow}}\left[ \frac{\mathbf{p}^2_{i\uparrow}}{2m} + \frac{1}{m} \mathbf{v}_{\uparrow}(\mathbf{x}_i) \cdot \mathbf{p}_{i\uparrow}  \right] \nonumber\\
 && + \sum_{j=1}^{N_{\downarrow}}\left[ \frac{\mathbf{p}^2_{j\downarrow}}{2m} + \frac{1}{m} \mathbf{v}_{\downarrow}(\mathbf{y}_j) \cdot \mathbf{p}_{j\downarrow} \right] 
 \nonumber\\
&& + \frac{1}{2m} \sum_{i=1}^{N_{\uparrow}} \left[\mathbf{v}^2_{\uparrow}(\mathbf{x}_i) + \xi_{\uparrow}(\mathbf{x}_i)  \right] \nonumber\\
&&+  \frac{1}{2m} \sum_{j=1}^{N_{\downarrow}} \left[ \mathbf{v}^2_{\downarrow}(\mathbf{y}_j) + \xi_{\downarrow}(\mathbf{y}_j) \right],
\end{eqnarray}
where the scalar potential $\xi_{\sigma}(\mathbf{r}) =   \hbar^2 \int d\mathbf{s}
 \left[\frac{\nabla_{\mathbf{r}} \chi_J(\mathbf{r}-\mathbf{s})}{\chi_J(\mathbf{r}-\mathbf{s})} \right]^2 
 \: \rho_{-\sigma}(\mathbf{s})$. This Hamiltonian describe a system of  CF interacting via \emph{short-range}
 vector $v_{\sigma}(\mathbf{r})$ and scalar $\xi_{\sigma}(\mathbf{r})$ potentials. The potentials depend on the density
 distribution of the opposite spin species. The CF Hamiltonian 
 still looks quite complicated, but it is well defined in the unitary limit where $|a_s| \to \infty$. 

   The above Hamiltonian, $H_\mathrm{CF}$, greatly simplifies upon performing a
 mean-field approximation. The approximation is similar in spirit to the mean-field theories 
 of composite particles in the fractional quantum Hall effect~\cite{Sondhi,Read,Jain}.
 Let us replace the density operator $\rho_{-\sigma}(\mathbf{s})$ of the CFs 
 by its expectation value in the expressions for $\mathbf{v}_{\sigma}(\mathbf{r})$ and 
 $\xi_{\sigma}(\mathbf{r})$. Thus, $ \mathbf{v}_{\sigma}(\mathbf{r}) = - i \hbar s_{\sigma} \int  d\mathbf{s}\, \ln \chi_J(\mathbf{r}-\mathbf{s})
\nabla_{\mathbf{s}}\rho_{-\sigma}(\mathbf{s})  = 0$
 and $\xi_{\sigma}(\mathbf{r}) = $ const. Dropping the constant term,  the mean-field 
 Hamiltonian reads:
 \begin{equation}
 H^\mathrm{MF}_\mathrm{CF} =  \sum_{i=1}^{N_{\uparrow}} \frac{\mathbf{p}^2_{i\uparrow}}{2m} 
 + \sum_{j=1}^{N_{\downarrow}} \frac{\mathbf{p}^2_{j\downarrow}}{2m}.
  \end{equation}  
In other words, the CFs become non-interacting and  their ground state wavefunction is 
the Slater determinant $\Phi_\mathrm{FS}(\mathbf{X},\mathbf{Y})$,
that is, the CF `Fermi sea'. The ground state for the actual fermions is $|\Psi_N \rangle = \mathcal{J}_{BPC}(a_s)|  \Phi_\mathrm{FS} \rangle$, 
which was introduced in Sect.~\ref{sec:projN} rather heuristically.   Furthermore,   
the excitations of this  state are obtained from projecting the excitations of  
non-interacting CF states, which are also Slater determinants.  In second quantization, these excitations 
correspond to the quasi-particles (created by  $\pi^{\dag}_{\sigma}(\mathbf{k})$ with $|\mathbf{k}| > k_{F\sigma}$, cf. Eq.~\ref{eq:qp1})
and the quasi-holes (created by $\eta_{\sigma}(\mathbf{k})$ with $|\mathbf{k}| < k_{F\sigma}$, cf. Eq.~\ref{eq:qh1}) introduced 
in Sect.~\ref{sec:projBCS}, and which were related to the Fermi polarons in Sect.~\ref{sec:projN}.  

  Nevertheless, it must be noted that the quasi-particle and quasi-hole states resulting from the projection  
are not necessarily orthogonal.  For single quasi-particle/quasi-hole states, spin and momentum conservation 
require that $\langle \Psi_N | \pi_{\sigma}(\mathbf{k}) \pi^{\dag}_{\sigma'}(\mathbf{p}) | \Psi_{N} \rangle = 0$ only
if $\mathbf{k} \neq \mathbf{p}$ and/or $\sigma \neq \sigma'$.  However, states containing different number of excitations but carrying the
same total momentum and spin are not  orthogonal. The non orthogonality stems from the non-unitary transformation
involving the operators $U_{\sigma}(\mathbf{r})$ introduced in Sect.~\ref{sec:projBCS}, which is needed to carry out the projection. 
 In spite of their lack of orthogonality, the  quasi-particle and  quasi-hole states are 
 left eigenstates of $\mathcal{J}_\mathrm{BPC}(a_s) H^\mathrm{MF}_\mathrm{CF}$
and therefore are stationary states at the mean-field level. However, because  of projection and 
the mean-field approximation,  $\mathcal{J}_\mathrm{BPC}(a_s)H^\mathrm{MF}_\mathrm{CF}$ is not a 
hermitian operator. This feature  makes it difficult to set up 
a perturbative approach to analyze the effect of corrections to the mean-field theory. Similar difficulties
have been found in the approach to the fractional Hall effect described in~\cite{Sondhi}. In 
standard Landau FLT, the quasi-particles are not  eigenstates of the system but narrow superpositions of the latter.
This leads to similar difficulties when defining quasi-particle operators in the standard theory of Fermi liquids~\cite{Nozieres}.
In the present  mean-field theory,  the matrix elements for quasi-particle decay vanish and therefore the quasi-particle 
and quasi-holes become stationary states. The non-orthogonality, however, can regarded as a remnant of the fact that they are not 
true eigenstates of the system.

 Although it may be coincidental as it results from a rather uncontrolled approximation, 
the lack of renormalization of the quasi-particle mass 
is in qualitative agreement with the experimental evidence around unitary. In this regard, 
the ENS group found that~\cite{ENS} $m^* \simeq 1.13 \: m$.  The lattice Monte Carlo calculations
of~\cite{Drut} are also in agreement with an internal energy that, up to an overall shift, is the same
as for a non-interacting Fermi gas.  If we rely on perturbation theory 
at weak coupling ($|k_F a_s| \ll 1$),  for $N_{\uparrow} = N_{\downarrow}$,  it is found that
$m^*/m  =  1+ \frac{8}{15 \pi^2} \left(7 \ln 2 -1 \right) (k_F a_s)^2 + O\left((k_F a_s)^2 \right)\simeq 1 +  0.208\: 
(k_F a_s)^2$ (see \emph{e.g.} Ref.~\cite{FetterW}, pag. 148). This formula yields about a $20 \%$ correction for $k_F a_s = -1$,
outside its validity range, and which would be about $10\%$ higher than the value of the effective mass determined by the ENS group
at unitarity.   In the highly polarized gas 
at unitarity  $m^*\simeq 1.17 \: m$~\cite{spinimbalance,ENS} and $m^*\simeq 1.20 \: m$ 
from the thermodynamic measuremnts of Ref.~\cite{ENS}. Furthermore, theoretical calculations of the Fermi polaron effective mass~\cite{Lobo,Combescot,Combescot2,Svistunov} also found that $m^*$ does not strongly renormalize from the bare atom mass  in a large region of 
the crossover regime where the normal phase is found to be stable.  Thus, phenomenologically, these small deviations from the bare mass
can be  thus regarded as measure of the  strength of the fluctuations  beyond the mean-field approximation.  The potential fluctuations 
can be related to those of the  operator $\nabla \rho_{\sigma}(\mathbf{r})$ within the range of the Jastrow factor $\sim \min\{|a_s|, k^{-1}_F \}$. 
In particular, when a molecular bound state forms, the mean-field theory will break down. This is because it assumes that the density around 
a given CF is constant and its fluctuations are negligible. But this is no longer true when the bound state forms. In other words, 
such a bound state cannot be described by a projecting a Slater determinant of CFs. 
Thus,  ultimately, the justification for the mean-field approximation 
is directly related to the accuracy of the 
description of the normal state by the Jastrow Slater wavefunctions. 
The latter have been shown to be good trial wavefunctions 
for the importance sampling in fixed-node diffusion Monte Carlo calculations~\cite{Giorgini,VMC,Pilati}, which
means that they are able to capture the nodal surface of the normal state.

 The thermodynamics of the normal state at the mean-field level is the same as for the 
 non-interacting Fermi gas because thermodynamic functions only depend on the spectrum
 and their degeneracies and not on the  wavefunctions.  Thus, Fermi liquid theory trivially applies to this system.
 To see this, we first recall that   the ground state $|\Psi_N\rangle$ has a well defined Fermi surface.
 Furthermore, the  quasi-particle and quasi-hole excitations are fermions that carry the same quantum numbers
 as the original fermions. This is because they result from projecting non-interacting Fermi gas
 states:
\begin{eqnarray}
\pi^{\dag}_{\sigma} (\mathbf{k}) |\Psi_N \rangle = \mathcal{J}_\mathrm{BPC}(a_s) \psi^{\dag}_{\sigma}(\mathbf{k}) |\Phi_\mathrm{FS} \rangle, \label{eq:qp1}\\
\eta_{\sigma} (\mathbf{k}) |\Psi_N \rangle = \mathcal{J}_\mathrm{BPC}(a_s) \psi_{\sigma}(\mathbf{k}) |\Phi_\mathrm{FS} \rangle.\label{eq:qh1}
\end{eqnarray}
In general, projection can be used as a formal device to establish a one to one correspondence between the states of the
 non-interacting gas of CFs  and the excitations of the real system. Just as in standard Landau FLT,  this allows to  introduce 
the distribution function of quasi-particles (or quasi-holes) $n_{\sigma}(\mathbf{k})$. The Landau free energy functional
within the mean-field approximation reads:
\begin{equation}
 F^{MF} =   F_{0} +  \sum_{\mathbf{k},\sigma} \left[\epsilon_{\sigma}(\mathbf{k}) - \mu_{\sigma} \right]  \: \delta n_{\sigma}(\mathbf{k})
\end{equation}
where $\delta n_{\sigma}(\mathbf{k}) = n_{\sigma}(\mathbf{k})-n^0_{\sigma}(\mathbf{k})$, and $n^0_{\sigma}(\mathbf{k}) = \theta(k_{F\sigma} - k)$
and $F_{0}$ is the ground state free energy. 
The Landau parameters vanish at the mean-field level.    Experimentally,  however, it is found that $F^s_{1} \neq 0$ ($F^s_{1} = 0.39$ for
the unitary unpolarized gas~\cite{ENS}), which by virtue of  Galilean invariance, accounts for the deviations of the effective mass from the bare mass. 
Furthermore, $F^s_0 = -0.42$ was determined by the ENS group~\cite{ENS} by fitting the FLT equation of state of the unpolarized gas at unitarity.
Thus, interactions between the composite Fermions which are described by the corrections to the mean-field theory should account for the Landau
parameters. 

 On the other hand, although the projection procedure  allows us to establish a one to one correspondence 
 with the non-interacting Fermi gas states,  it 
does not quite correspond to  the process of adiabatic continuity envisaged 
by Landau (see \emph{e.g.} Ref.~\cite{Nozieres}, pag. 2 and following). In the case of $^{3}$He, the Jastrow Slater wavefunctions
describe  a standard
Fermi liquid. 
However, in this case, unlike  the case of $^{3}$He  where the Jastrow factor is introduced to obtain 
a better variational description of the ground state, the Jastrow factor introduced above simply
implements the BPC, that is, the projection onto the physical Hilbert space. It is therefore not
just a  convenience, but a physical need. 
Thus,  when dealing with the  quasi-particle and quasi-hole excitations, we are effectively 
projecting non-interacting states~\cite{Anderson} (cf. Eqs.~\ref{eq:qp1},\ref{eq:qh1} ). It can be argued that,
by using the device of  varying the scattering length so that $a_s \to 0$, the correspondence with the particle and hole states of
the non-interacting Fermi gas  becomes more apparent.  Such a procedure does not warranty that the overlaps:
\begin{eqnarray}
Z_\mathrm{qp} &=& \lim_{k \to k_F^{+}}\langle  \Psi_\mathrm{N}  | \psi_{\sigma}(\mathbf{k})  \pi^{\dag}_{\sigma}(\mathbf{k})  | \Psi_\mathrm{N} \rangle, \\ 
Z_\mathrm{qh} &=& \lim_{k \to k_F^{-}} \langle  \Psi_\mathrm{N}  | \psi^{\dag}_{\sigma}(\mathbf{k}) \eta_{\sigma}(\mathbf{k})  | \Psi_\mathrm{N} \rangle 
\end{eqnarray}
are in general equal, as required by \emph{standard} FLT theory,  
or even remain finite for all values of $a_s$ in the thermodynamic limit. 
If they vanished, the quasi-particles and quasi-holes, which are the  true elementary
excitations of the system, will not behave as  
Landau quasi-particles~\cite{Anderson,AndersonJain}. 
The deep reason why $Z_\mathrm{qp}$ and $Z_\mathrm{qh}$ may turn out to be different or even vanish is the following:
When adding one fermion to the ground state $|\Psi_N\rangle$ (or in general to any physical state),
the resulting state is not in the physical Hilbert space. This is  because the other particles cannot
immediately adapt to the newcomer. However, this conclusion does not apply to the state that
results from removing one fermion. Such a state  belongs  the physical space but also contains the
'correlation hole' left by the removed particle. The latter situation is relevant to spectroscopic measurements,
such as those carried out by the JILA group~\cite{JILA}. Therefore, there is a clear asymmetry between the states with one particle 
more and one particle less.  Investigating these issues is beyond the scope of the present work and 
will be done elsewhere~\cite{inprogress}.

\section{Conclussions}\label{sec:concl}
  
  In previous sections, we have described a wavefunction-based approach to study the properties of the
 ultracold dilute Fermi gas. The basic idea is that projection onto the physical Hilbert space of wave functions
 obeying the Bethe-Peierls condition must be implemented from the scratch.
The need for projection stems from the short range of the interaction and
 diluteness of the gas. This implies that two-body encounters are described
by the  two-body wavefunction, which is assumed to reach its asymptotic
value before other collisions can take place. This physics is encapsulated in the Bethe-Peiers
condition, which in this work has been  implemented by means of a Jastrow factor.  It is 
interesting to note  that the situation is reminiscent of the composite Fermion approach of the
fractional quantum Hall effect~\cite{Read,Jain,Sondhi}. It 
is also related to  Anderson's `hidden Fermi liquid  theory' ~\cite{Jain,Anderson} of the
Gutzwiller projected Hubbard model. The common thread between these ideas is that appropriate projection of 
a Fermi gas defined in an unphysical Hilbert space can account for Fermi-liquid like 
behavior in systems with no obvious small parameter for a perturbative analysis to be carried out. 
This is the case of the ultracold dilute Fermi gas in the crossover regime. 
 
 Furthermore, as far as the superfluid state is concerned, in Sect.~\ref{sec:projBCS}
we have argued that  projection automatically fixes an important 
shortcoming of the BCS wavefunction pointed by Tan~\cite{Tan}.
By considering  that the gap of the (unprojected) BCS state collapses,
the same principle leads to a projected Fermi gas, which is postulated as the most natural
candidate for the normal state of the system, even in the strongly interacting limit. 
This state is described by Jastrow-Slater wavefunction, which is also shown to be
the ground state of a  mean-field theory. The mean field theory also
exhibits non-interacting  quasi-particle and quasi-hole excitations, which at low temperatures yield
Fermi liquid thermodynamics. The quasi-particles are related to the Fermi polarons,
which have been found to describe the normal state of a highly polarized Fermi gas 
at unitarity~\cite{Lobo,spinimbalance}. We have also shown that the superfluid ground
state can be understood as a condensate of pairs of composite Fermions or Fermi polarons.

 A rather direct way of approaching the composite fermion theory is to 
investigate numericaly the properties low-lying excited states of  the normal and superfluid state state. 
This has been done already to a large extent for  ground state properties
using the fixed-node diffusion Monte Carlo method~\cite{Giorgini,VMC,Pilati}. However, it is also possible to
employ the same Jastrow projector to analyze  excitations of the system as in Jain's approach to the theory of
composite fermions of the fractional quantum Hall effect~\cite{Jain}.
For example,  a better understanding of the quasi-particle interactions can be obtained through a Monte Carlo 
calculation of the energies of two quasi-particle (two quasi-hole, o quasi-particle and quasi-hole) excitations, which will allow to
estimate the Landau parameters. Furthermore, analytical progress may be also possible
by studying corrections to the mean-field approach discussed above. However, it may be 
necessary to find more convenient ways of implementing the projection, \emph{e.g.}  
by  defining a constrained  functional integral. This would make it possible to develop a field
theoretical approach to deal with the composite fermion or Fermi polaron liquids, and systematically 
study corrections to the mean-field theory. As argued in previous
sections, we expect the composite fermions to be much weakly interacting than the original fermionic atoms, 
as  part of the interactions 
have been  transfered to the Jastrow factor.   If successful, such efforts will take the theory from the present 
qualitative level to one where quantitative predictions can be made and 
compared directly with the experiments.  Hopefully, progress on this problem 
will be reported elsewhere~\cite{inprogress}. 

\acknowledgments
 I thank P. Zhang for movitaving my interest in this problem and 
 for  bringing to my attention Tan's observation  about the BCS wavefunction, P. Naidon for 
 discussions and comments on an early version of the manuscript, 
 and T. Eggel for stimulating discussions. Discussions with M. Ueda, and G. Vignale  are 
 also gratefully acknowledged.   I am also grateful to T. L. Ho for his valuable advice to improve the presentation 
 and M. Oshikawa for his kind hospitality  at the Institute for Solid State Physics 
(University of Tokyo) .

\appendix


\begin{thebibliography}{30}
%
\bibitem{Duke}
J. Kinast, S.L. Hemmer, M. Gehm, A. Turlapov, J.E. Thomas, Phys. Rev. Lett. {\bf 92}, 150402 (2004);
L. Luo \emph{et al.} \emph{ibid}. {\bf 98}, 080403 (2007);
L. Luo and J. Thomas, J. Low Temp. Phys. {\bf 154}, 1 (2009).
%
\bibitem{ENS}
S. Nascimb\`ene, N. Navon, K. J. Jiang, F. Chevy, and C. Salomon, Nature {\bf 463}, 1057 (2010). 
%
\bibitem{Tokyo}
M. Horikoshi, S. Nakajima, M. Ueda, and T. Mukaiyama,  Science {\bf 327}, 442 (2010).
%
\bibitem{Ho_universal}
T. L. Ho, Phys. Rev. Lett. {\bf 92}, 090402 (2004).
%
\bibitem{JILA}
J. ~P. Gaebler, J.~T. Stewart, T.~E. Drake, D.~S. Jin, A. Perali, P. Pieri, and 
G.~C. Strinati, arXiv:1003.1147 (2010).
%
\bibitem{ARPES}
T.-L. Dao, A. Georges, J. Dalibard, C. Salomon, I. Carusotto,  Phys. Rev. Lett. {\bf 98}, 240402 (2007);
I. Stewart, J. P. Gaebler, and D. S. Jin, Nature {\bf 454}, 744 (2008). 
%
\bibitem{pseudogapth}
C.-C. Chien, H. Guo, and K. Levin, Phys. Rev. A {\bf 81}, 023622 (2010);
Q.  Chen and K. Levin, Phys. Rev. Lett. {\bf 102}, 1904020 (2009);
Q. Chen, J.  Stajic, S. Tan, and K. Levin, Phys. Rep. {\bf 412}, 1 (2005).
P. Pieri, A. Perali, G.~C. Strinati, Nature Physics 5, 736-740 (2009),
and references therein.
%
\bibitem{Randeriaps}
See M. Randeria, cond-mat/9710223 (1997) for a review. 
%
\bibitem{Tsuchiya}
S. Tsuchiya,  R. Watanabe, Y. Ohashi, Phys. Rev. A {\bf 80} 033613 (2009);
%
\bibitem{Punk}
M. Punk and W. Zwerger, Phys. Rev. A {\bf 063612} (2009).
%
\bibitem{Lobo}
C. Lobo, A. Recati, S Giorgini, and S. Stringari, Phys.~Rev.~Lett {\bf 97}, 200403 (2006).
%
\bibitem{spinimbalance}
A. Schirotzek, C.-H. Wu, A. Sommer, and M. W. Zwierlein, Phys. Rev. Lett. {\bf 102} 230402 (2009);
%
\bibitem{ENS2}
S. Nascimb\`ene, N. Navon, K. J. Jiang, L. Tarruell, M. Teichmann, J. McKeever, F. Chevy, and C. Salomon
Phys. Rev. Lett. {\bf 103}, 170402 (2009). 
%
\bibitem{Drummond}
P. Magierski, G. Wlazlowski, A. Bulgac, J. Drut, Phys. Rev. Lett. {\bf 103}, 210403 (2009); 
H. Hu, X.-J. Liu,  P. Drummond, and H. Dong, arxiv:1003.1538 (2010).
%
\bibitem{Sheehy}
M. Y. Veillette, D. E. Sheehy, and L. Radzihovsky,  Phys.Rev. A {\bf 75},  043614 (2007).
%
\bibitem{Sheehy2}
M. Veillette, E. G. Moon, A. Lamacraft, L. Radzihovsky, S. Sachdev, and D.E. Sheehy, 
Physical Review A {\bf 78}, 033614 (2008) 
%
\bibitem{Nikolic}
P. Nikolic and S. Sachdev, Phys. Rev. A {\bf 75},  033608 (2007).
%
\bibitem{Son}
Y. Nishida and D. T. Son, Phys.Rev.Lett. {\bf 97}  050403 (2006);
Phys. Rev. A {\bf 75}, 063617 (2007);  arXiv:1004.3597 (2010).
%
\bibitem{Svistunov}
N. Prokof'ev and B. Svistunov, Phys. Rev. B {\bf 77}, 020408 (2008);
Phys. Rev. B 77, 125101 (2008).
%
\bibitem{Loram}
J. W. Loram \emph{et al.} Phys. Rev. Lett. {\bf 71}, 1740 (1993). 
%
\bibitem{Drut}
A. Bulgac, J. Drut, and P. Magierski,   Phys. Rev. A {\bf 78},  023625 (2008).
%
\bibitem{Tan}
S. Tan, Adv. in Phys. (NY) {\bf 323}, 2971 (2008);
Annals of Physics {\bf 323} 2952  (2008). 
%
\bibitem{JILA_Tan}
J.~T. Stewart, J.~P. Gaebler, T.~E. Drake, and D.~S. Jin,  arXiv:1002.1987 (2010).
%
\bibitem{LeggettBCSBEC}
D. M. Eagles, Phys. Rev. {\bf 186}, 456 (1969);
A. J. Leggett in \emph{Modern Trends in the Theory of Condensed Matter}. Edited
by A. Pekalski and R. Przystawa (Springer-Verlag, Berlin, 1980);
C.~A.~R. S\'a de Melo, M. Randeria, and J. R. Engelbrecht
Phys. Rev. Lett. 71, 3202 (1993). 
%
\bibitem{Trento}
S. Giorgini, S. Stringari, L. P. Pitaevskii, Rev. Mod. Phys. {\bf 80}, 1215 (2008). 
%
\bibitem{Castin}
Y. Castin in 
\emph{Proceedings of the Enrico Fermi Varenna School on Fermi gases} (2006), 
Ultra-cold Fermi Gases, M. Inguscio, W. Ketterle, C. Salomon (Ed.) (2007) 289-349;
arxiv:0612613 (2006). 
%
\bibitem{LeeYangHuang}
K. Huang and C.~N. Yang, Phys. Rev. {\bf 105}, 767 (1957); 
T.~D. Lee and C. N. Yang, \emph{ibid} {\bf 105},  1119 (1957);
T.~D. Lee, K. Huang, and C. N. Yang, \emph{ibid} {\bf 106}, 1135 (1957). 
%
\bibitem{MoraWaintal}
C. Mora and X. Waintal, Phys. Rev. Lett. {\bf 99}, 030403 (2007).
%
\bibitem{Chevy}
F. Chevy, Phys. Rev.  A {\bf 74}, 063628 (2006).
F. Chevy and C. Mora, arxiv:1003.080 (2010).
%
\bibitem{Combescot}
R. Combescot and S. Giraud, Phys. Rev. Lett. {\bf 101}, 050404 (2008).
%
\bibitem{Punk2}
M. Punk, P. T. Dumitrescu, and W. Zwerger, Phys. Rev. A {\bf 80}, 053605 (2009).
%
\bibitem{Combescot2}
R. Combescot, A. Recati, C. Lobo, and F. Chevy,
Phys. Rev. Lett. {\bf 98}, 180402 (2007).
%
\bibitem{VMC}
J. Carlson, S.-Y. Chang, V. R. Pandharipande, and K.~E. Schmidt, Phys. Rev. Lett. {\bf 91}, 050401 (2003). 
G.~E. Astrakharchik, J. Boronat, J. Casulleras, and S. Giorgini,
Phys. Rev. Lett. 95, 230405 (2005);
A.~J. Morris, P. L\'opez-Rios, R.~J. Needs, Phys. Rev. A {\bf 81}, 033619 (2010). 
%
\bibitem{Bouchaudetal}
J.~P. Bouchaud, A. Georges, and C. Lhuillier, J. Phys. (Paris) {\bf 49}, 553 (1988). 
%
\bibitem{Nozieres}
P. Nozi\`eres, \emph{Theory of Interacting Fermi systems},  Perseus books (1997).
%
\bibitem{Jain}
J. K. Jain, Phys. Rev. Lett. {\bf 63}, 199 (1989);
J. K. Jain, \emph{Composite Fermions}, Cambridge University Press (Cambridge UK, 2007).
%
\bibitem{Read}
N. Read, Phys. Rev. Lett. {\bf 62}, 86 (1988).
%
\bibitem{Sondhi}
R. Rajaraman and S. Sondhi, Int. J. Mod. Phys. B {\bf 10}, 793 (1996); 
R. Rajaraman, Phys. Rev. B {\bf 56}, 6788 (1997);
Y. Yue and Y.~S. Wu, arxiv:9608061 (1996).
%
\bibitem{Giorgini}
G.~E. Astrakharchik, J. Boronat, J. Casulleras, and S. Giorgini,
 Phys. Rev. Lett. {\bf 93}, 2000404 (2004).
%
\bibitem{Pilati}
S. Pilati and S. Giorgini, Phys. Rev. Lett. {\bf 100}, 030401 (2008). 
%
\bibitem{FetterW}
A. Fetter and J.~K. Wallecka, \emph{Quantum Theory of Many-Particle Physics}
Dover Publications (New York, 2003).
%
\bibitem{Anderson}
P. W. Anderson,  Phys. Rev. B {\bf 78}, 174505 (2008).
%
\bibitem{AndersonJain}
J. K. Jain and P. W. Anderson Proc. Natl. Acad. Sci. (U.S.A) {\bf 106}, 9131 (2009). 
%
\bibitem{inprogress}
M. A. Cazalilla, work in progress.
 \end{thebibliography}
\end{document}